\documentclass[aps,showpacs,preprintnumbers,amsmath, amssymb]{revtex4}

\usepackage{float}
\usepackage{graphics,epsfig}
\usepackage{graphicx}
\usepackage{dcolumn}
\usepackage{bm}

\begin{document}
\baselineskip=0.8 cm
\title{{\bf Analytical investigation of the phase transition between holographic insulator and superconductor in Gauss-Bonnet gravity}}

\author{Qiyuan Pan$^{1,2}$\footnote{panqiyuan@126.com}, Jiliang Jing$^{1,2}$\footnote{jljing@hunnu.edu.cn}, Bin Wang$^{3}$\footnote{wang~b@sjtu.edu.cn}}
\affiliation{$^{1}$Institute of Physics and Department of Physics,
Hunan Normal University, Changsha, Hunan 410081, China}
\affiliation{$^{2}$ Key Laboratory of Low Dimensional Quantum
Structures and Quantum Control of Ministry of Education, Hunan
Normal University, Changsha, Hunan 410081, China}
\affiliation{$^{3}$ INPAC and Department of Physics, Shanghai Jiao
Tong University, Shanghai 200240, China}

\vspace*{0.2cm}
\begin{abstract}
\baselineskip=0.6 cm
\begin{center}
{\bf Abstract}
\end{center}

We employ the variational method for the Sturm-Liouville eigenvalue
problem to analytically study the phase transition between the
holographic insulator and superconductor in the Gauss-Bonnet
gravity. By investigating the s-wave and p-wave holographic
insulator/superconductor models, we find that this analytic method
is more effective to obtain the analytic results on the condensation
and the critical phenomena in the AdS soliton background in
Gauss-Bonnet gravity. Our analytic result can be used to back up the
numerical computations in the AdS soliton with Gauss-Bonnet
correction.

\end{abstract}
\maketitle
\newpage
\vspace*{0.2cm}

\section{Introduction}

The profound finding of the anti-de Sitter/conformal field theory
(AdS/CFT) correspondence \cite{Witten,Maldacena,Gubser1998} has
provided a framework to describe the strongly coupled field theories
in a weakly coupled gravitational system. Recently, this
correspondence has been applied to study the holographic model of
superconductors in which a remarkable connection has been observed
between the gravitational physics and the condensed matter physics
\cite{GubserPRD78}. It has been shown that the bulk AdS black hole
becomes unstable and scalar hair condenses below a critical
temperature. The instability of the bulk black hole corresponds to a
second order phase transition from normal state to superconducting
state which brings the spontaneous U(1) symmetry breaking. In the
boundary dual CFT, these properties exhibit the behavior of the
superconductor \cite{HartnollPRL101,HartnollJHEP12}. Due to the
potential applications to the condensed matter physics, the
condensation in bulk AdS black holes has been investigated
extensively, for reviews, see Refs.
\cite{HartnollRev,HerzogRev,HorowitzRev} and references therein.

In additional to the bulk AdS black hole spacetime, recently it was
found that a holographic model can be constructed in the bulk AdS
soliton background to describe the insulator and superconductor
phase transition \cite{Nishioka-Ryu-Takayanagi}. Adding the chemical
potential to the AdS soliton, a second order phase transition can
happen when the chemical potential is over a critical value. This
phase transition can be used to describe the transition between the
insulator and superconductor, while it is different from the
Hawking-Page phase transition between the Ricci flat AdS black hole
and the AdS soliton \cite{Surya-Schleich}. Taking the backreaction
of the matter fields into account, it was argued that  the order of
the phase transition can be changed from the second to the first if
the backreaction is strong enough \cite{Horowitz-Soliton}. In the
St$\ddot{u}$ckelberg mechanism, rich physics on the phase transition
between the holographic insulator and superconductor in the AdS
soliton background has been observed \cite{peng-pan-wang}. Further
investigations on various insulator and superconductor phase
transitions in different theories of gravity have been carried out
\cite{Akhavan-Soliton,Basu-Soliton,brihaye-Soliton,Cai-Li-Zhang,Pan-Wang}.

In most cases, the holographic superconductors were studied
numerically. Ideally, one would like to have a full analytic
description of the phase transition and condensation phenomena. In
addition to back up numerical results, the analytic description can
help to gain more insights, for example it may tell what properties
of the action decide the mean-field behaviors etc. Recently, there
appeared two analytic approaches in parallel to the numerical
calculation. One is the analytic matching method which was first
proposed in \cite{Gregory} and later refined in
\cite{Pan-Wang,KannoCQG}. With this method, we can calculate the
critical temperature analytically within a few percent in the best
case. This analytic approach has been extended to derive the upper
critical magnetic field when the holographic superconductor is
immersed in constant external magnetic fields \cite{Ge-Wang,Ge2011}.
In higher dimensions, this analytic method can keep valid only when
the matching point is chosen within an appropriate range
\cite{Pan-Wang}. However the matching method is not effective to
describe the AdS soliton, neither can it be used to derive the
critical exponents for the condensation near the critical
temperature. The mean-field critical exponent $1/2$ at the critical
temperature comes mostly from numerically solving the holographic
systems and doing data fitting. In \cite{Siopsis,SiopsisBF}, the
authors extended the variational method for the Sturm-Liouville
(S-L) eigenvalue problem to analytically calculate the critical
exponent near the critical temperature. This method was further
applied to analytically study some properties of holographic
superconductors in AdS black hole backgrounds in Einstein gravity
\cite{ZengSL} and Einstein-Gauss-Bonnet gravity \cite{Li-Cai-Zhang}
in the probe limit, respectively.

An analytic study by using the S-L method on the phase transition
between the holographic insulator and superconductor was done in
\cite{Cai-Li-Zhang} in the Einstein gravity. It is of interest to
further generalize the S-L method to study holographic
superconductor developed in the AdS soliton background in the
Gauss-Bonnet gravity. The condensation phenomena and the phase
transition between the s-wave holographic insulator and
superconductor in the Gauss-Bonnet gravity were investigated
numerically in \cite{Pan-Wang,Liu-Wang}. We will also extend the
investigation of the p-wave holographic insulator and superconductor
phase transition with Gauss-Bonnet correction in this work, which
has not been constructed as far as we know. It is not trivial to
analytically study the condensation and the phase transition by
taking into account of the influence of the Gauss-Bonnet coupling.
Besides to be used to check  numerical computation, the analytic
investigation can clearly disclose the critical exponent of the
system at the critical temperature and the influence of the
Gauss-Bonnet factor in the phase transition. In the AdS soliton
background, we will compare two available analytic methods and argue
that the S-L method is more effective for the analytic study of the
condensation.

The plan of the work is the following. In Sec. II we briefly review
the AdS soliton background in the Gauss-Bonnet gravity. In Sec. III
we explore the s-wave insulator and superconductor phase transition
with Gauss-Bonnet correction. In Sec. IV we discuss the p-wave case.
We conclude in the last section with our main results.

\section{Gauss-Bonnet AdS soliton}

In order to study the superconducting phase dual to the Guass-Bonnet AdS
soliton configuration in the probe limit, we start with the
five-dimensional AdS soliton in the Gauss-Bonnet gravity in the form
\cite{Cai-Kim-Wang}
\begin{eqnarray}\label{soliton}
ds^2=-r^{2}dt^{2}+\frac{dr^2}{f(r)}+f(r)d\varphi^2+r^{2}(dx^{2}+dy^{2}),
\end{eqnarray}
with
\begin{eqnarray}
f(r)=\frac{r^2}{2\alpha}\left[1-\sqrt{1-\frac{4\alpha}{L^{2}}
\left(1-\frac{r_{s}^{4}}{r^{4}}\right)}~\right],
\end{eqnarray}
where $r_{s}$ is the tip of the soliton which is a conical
singularity in this solution, $\alpha$ is the Gauss-Bonnet coupling
constant and $L$ is the AdS radius. It should be noted that in the
asymptotic region ($r\rightarrow\infty$), we find
\begin{eqnarray}
f(r)\sim\frac{r^2}{2\alpha}\left(1-\sqrt{1-\frac{4\alpha}{L^2}}
\right)\,,
\end{eqnarray}
so the effective asymptotic AdS scale can be defined by
\cite{Cai-2002,Cai-Kim-Wang}
\begin{eqnarray}\label{LeffAdS}
L^2_{\rm eff}=\frac{2\alpha}{1-\sqrt{1-\frac{4\alpha}{L^2}}} \to
\left\{
\begin{array}{rl}
L^2   \ , &  \quad {\rm for} \ \alpha \rightarrow 0~, \\
\frac{L^2}{2}  \ , &  \quad {\rm for} \  \alpha \rightarrow
\frac{L^2}{4}~.
\end{array}\right.
\end{eqnarray}
$\alpha=L^{2}/4$ is the Chern-Simons limit corresponding to the
upper bound of the Gauss-Bonnet factor. When $\alpha\rightarrow0$, (\ref{soliton})
goes back to the Schwarzschild AdS soliton. For simplicity, in the following we will consider the Gauss-Bonnet factor in the range $0<\alpha\leq
L^{2}/5$. For the smoothness at the tip, we impose a
period $\beta=\frac{4\pi L^{2}}{(d-1)r_{s}}$ for the coordinate
$\varphi$ to remove the singularity.

\section{Phase transition between the s-wave insulator and superconductor}

In the background of the Gauss-Bonnet-AdS soliton, we consider a
Maxwell field and a charged complex scalar field coupled via the
action
\begin{eqnarray}\label{System}
S=\int d^{5}x\sqrt{-g}\left[
-\frac{1}{4}F_{\mu\nu}F^{\mu\nu}-|\nabla\psi - iA\psi|^{2}
-m^2|\psi|^2 \right] \ .
\end{eqnarray}
Taking the ansatz of the matter fields as $\psi=\psi(r)$ and
$A=\phi(r) dt$, we can get the equations of motion for the scalar
field $\psi$ and gauge field $\phi$ in the form
\begin{eqnarray}
\psi^{\prime\prime}+\left(
\frac{f^\prime}{f}+\frac{3}{r}\right)\psi^\prime
+\left(\frac{\phi^2}{r^2f}-\frac{m^2}{f}\right)\psi=0\,, \label{Psi}
\end{eqnarray}
\begin{eqnarray}
\phi^{\prime\prime}+\left(\frac{f^\prime}{f}+\frac{1}{r}\right)
\phi^\prime-\frac{2\psi^2}{f}\phi=0, \label{Phi}
\end{eqnarray}
where the prime denotes the derivative with respect to $r$.

In order to solve the above equations, we have to impose the
boundary conditions at the tip $r=r_{s}$ and at
$r\rightarrow\infty$. At the tip $r=r_{s}$, the solutions behave as
\begin{eqnarray}
\psi=\tilde{\psi}_{0}+\tilde{\psi}_{1}(r-r_{s})+\tilde{\psi}_{2}(r-r_{s})^{2}+\cdots\,, \nonumber \\
\phi=\tilde{\phi}_{0}+\tilde{\phi}_{1}(r-r_{s})+\tilde{\phi}_{2}(r-r_{s})^{2}+\cdots\,,
\label{SolitonBoundary}
\end{eqnarray}
where $\tilde{\psi}_{i}$ and $\tilde{\phi}_{i}$ ($i=0,1,2,\cdots$)
are integration constants, and the Neumann-like boundary
condition has been imposed to keep every physical quantity finite
\cite{Nishioka-Ryu-Takayanagi}. It is worth noticing that one can
find a constant nonzero gauge field $\phi(r_{s})$ at $r=r_{s}$, in
contrary to that of the AdS black hole where $\phi(r_{+})=0$ at the
horizon.

Near the boundary $r\rightarrow\infty$, we have asymptotic
behaviors
\begin{eqnarray}
\psi=\frac{\psi_{-}}{r^{\lambda_{-}}}+\frac{\psi_{+}}{r^{\lambda_{+}}}\,,\hspace{0.5cm}
\phi=\mu-\frac{\rho}{r^{2}}\,, \label{infinity}
\end{eqnarray}
where $\mu$ and $\rho$ are interpreted as the chemical potential and
charge density in the dual field theory respectively. Here
$\lambda_\pm=2\pm\sqrt{2+m^{2}L_{\rm eff}^2}$. The coefficients
$\psi_{-}$ and $\psi_{+}$ both multiply normalizable modes of the
scalar field equations and they correspond to the vacuum expectation
values $\psi_{-}=<\mathcal{O}_{-}>$, $\psi_{+}=<\mathcal{O}_{+}>$ of operators dual to the scalar field according to the AdS/CFT correspondence. We can impose boundary conditions that
either $\psi_{-}$ or $\psi_{+}$ vanish
\cite{HartnollPRL101,HartnollJHEP12}. For simplicity, we will scale
$L=1$ and $r_{s}=1$ in the following just as in
\cite{Nishioka-Ryu-Takayanagi,Pan-Wang}.

Before going further, we would like to give a comment. In the AdS
soliton background, since at the tip $\phi(r_{s})$ does not vanish
in (6) and (7), which leads that the ${\phi}^2$ terms in the coupled
equations cannot be got rid of as did in the AdS black hole case, we
cannot count on the matching method to obtain the analytic result.
Here we will apply the S-L method \cite{Siopsis} to analytically
investigate the properties of the s-wave holographic
insulator/superconductor phase transition in the Gauss-Bonnet
gravity. We will calculate the critical chemical potential to
accommodate the phase transition and analytically derive  the
critical exponent of condensation operator. In addition, we will
derive the relation between the charge density and the chemical
potential near the phase transition point and examine the effect of
the Gauss-Bonnet factor.

\subsection{Critical chemical potential}

Introducing a new variable $z=r_{s}/r$, we can rewrite Eqs.
(\ref{Psi}) and (\ref{Phi}) into
\begin{eqnarray}
\psi^{\prime\prime}+\left(
\frac{f^\prime}{f}-\frac{1}{z}\right)\psi^\prime
+\left(\frac{\phi^2}{z^2f}-\frac{m^2}{z^4f}\right)\psi=0\,,
\label{Psi-Z}
\end{eqnarray}
\begin{eqnarray}
\phi^{\prime\prime}+\left(\frac{f^\prime}{f}+\frac{1}{z}\right)
\phi^\prime-\frac{2\psi^2}{z^4f}\phi=0, \label{Phi-Z}
\end{eqnarray}
where the prime here denotes the derivative with respective to $z$.

It has been shown numerically that the solution is unstable and a
hair can be developed when the chemical potential is bigger than a
critical value, i.e., $\mu>\mu_{c}$. For lower chemical potential,
$\mu<\mu_{c}$, the gravitational dual is an AdS soliton with a
nonvanishing profile for the scalar field $\psi$, which can be
viewed as an insulator phase
\cite{Nishioka-Ryu-Takayanagi,Pan-Wang}. Thus, there is a phase
transition between the insulator and superconductor phases around
the critical chemical potential $\mu_{c}$.

At the critical chemical potential $\mu_{c}$, the scalar field
$\psi=0$. So near the critical point Eq. (\ref{Phi-Z}) reduces to
\begin{eqnarray}
\phi^{\prime\prime}+\left(\frac{f^\prime}{f}+\frac{1}{z}\right)\phi^\prime=0.
\label{Phi-critical}
\end{eqnarray}
With the Neumann-like boundary condition (\ref{SolitonBoundary}) for
the gauge field $\phi$ at the tip $r=r_{s}$, we can obtain the
physical solution $\phi(z)=\mu$ to Eq. (\ref{Phi-critical}) when
$\mu<\mu_{c}$. Considering the asymptotic behavior in Eq.
(\ref{infinity}), close to the critical point this solution
indicates that $\rho=0$ near the AdS boundary $z=0$, which gives
fairly good agreement with numerical results in Ref.
\cite{Pan-Wang}.

As $\mu\rightarrow\mu_{c}$, the scalar field equation (\ref{Psi-Z})
reduces to
\begin{eqnarray}
\psi^{\prime\prime}+\left(\frac{f^\prime}{f}-\frac{1}{z}\right)\psi^\prime
+\left(\frac{\mu^2}{z^2f}-\frac{m^2}{z^4f}\right)\psi=0.
\label{criticalPsi}
\end{eqnarray}
As in \cite{Siopsis}, we introduce a trial function $F(z)$ near
the boundary $z=0$ which satisfies
\begin{eqnarray}\label{PhiFz}
\psi(z)\sim \langle{\cal O}_{i}\rangle z^{\lambda_i}F(z),
\end{eqnarray}
with $i=+$ or $i=-$. Here we will impose the boundary condition
$F(0)=1$ and $F'(0)=0$. Then, we can obtain the equation of motion
for $F(z)$
\begin{eqnarray}\label{Fzmotion}
F^{\prime\prime}+\left[\frac{2\lambda_i}{z}+\left(\frac{f'}{f}-\frac{1}{z}\right)\right]
F^{\prime}+\left[\frac{\lambda_i(\lambda_{i}-1)}{z^2}+\frac{\lambda_i}{z}
\left(\frac{f'}{f}-\frac{1}{z}\right)+\frac{1}{z^{4}f}(\mu^{2}z^{2}-m^{2})\right]F=0.
\end{eqnarray}
Defining a new function
\begin{eqnarray}
T(z)=\frac{z^{2\lambda_{i}-3}\sqrt{1+4(z^{4}-1)\alpha}-1}{2\sqrt{\alpha}},
\end{eqnarray}
we can rewrite Eq. (\ref{Fzmotion}) as
\begin{eqnarray}\label{NewFzmotion}
(TF^{\prime})^{\prime}+T\left[\frac{\lambda_i(\lambda_{i}-1)}{z^2}+\frac{\lambda_i}{z}
\left(\frac{f'}{f}-\frac{1}{z}\right)+\frac{1}{z^{4}f}(\mu^{2}z^{2}-m^{2})\right]F=0.
\end{eqnarray}
According to the Sturm-Liouville eigenvlaue problem
\cite{Gelfand-Fomin}, we obtain the expression which can be used to
estimate the minimum eigenvalue of $\mu^2$
\begin{eqnarray}\label{eigenvalue}
\mu^{2}=\frac{\int^{1}_{0}T\left(F'^{2}-UF^{2}\right)dz}{\int^{1}_{0}VF^{2}dz},
\end{eqnarray}
with
\begin{eqnarray}
&&U=\frac{\lambda_{i}(\lambda_{i}-1)}{z^{2}}+\frac{\lambda_{i}}{z}\left(\frac{f'}{f}-\frac{1}{z}\right)
-\frac{m^{2}}{z^{4}f},\nonumber\\
&&V=\frac{T}{z^{2}f}.
\end{eqnarray}
In the following calculation, we will assume the trial function to
be $F(z)=1-az^{2}$, where $a$ is a constant.

In \cite{Pan-Wang} the condensate $\langle{\cal O}_{+}\rangle$ was
numerically calculated in 5-dimensional Gauss-Bonnet AdS soliton
background simply by fixing $\psi_{-}=0$. It has been shown numerically that the
increase of the Gauss-Bonnet factor $\alpha$ results in the increase of the critical
chemical potential, which means that the higher curvature correction will make it
harder for the scalar hair to be condensated. Now we can use the
S-L method to understand the condensation analytically.

Using Eq. (\ref{eigenvalue}) to compute the minimum eigenvalue of
$\mu^{2}$ for $i=+$, we can obtain the critical chemical potential
$\mu_{c}$ for different strength of the curvature correction and the mass of the scalar field.
As an example, we calculate the case of $m^{2}L_{\rm eff}^2=0$ in
detail. From Eq. (\ref{eigenvalue}), we obtain
\begin{eqnarray}
\mu^{2}=\frac{\Sigma(a,\alpha)}{\Xi(a,\alpha)},
\end{eqnarray}
with
\begin{eqnarray}
\Sigma(a,\alpha)&=&\frac{1-6\alpha+\sqrt{1-4\alpha}(-1+4\alpha)}{3\alpha^{3/2}}+
\frac{1}{16\alpha^{2}}\left[2\sqrt{\alpha}(-3+20\alpha)+3(1-4\alpha)^{2}\log\frac{\sqrt{\alpha}+2\alpha}{\sqrt{\alpha(1-4\alpha}})\right]a
\nonumber\\
&&+\frac{3[-1+10\alpha-30\alpha^{2}+\sqrt{1-4\alpha}(1-8\alpha+16\alpha^{2})]}{40\alpha^{5/2}}a^{2},
\nonumber\\
\Xi(a,\alpha)&=&\frac{(-10+15a-6a^{2})\sqrt{\alpha}}{60}.
\end{eqnarray}
For different values of the Gauss-Bonnet factor, we can get the
minimum eigenvalues of $\mu^{2}$ and the corresponding values of
$a$, for example, $\mu_{min}^{2}=11.607$ and $a=0.440$ for
$\alpha=0.0001$, $\mu_{min}^{2}=12.667$ and $a=0.386$ for
$\alpha=0.1$ and $\mu_{min}^{2}=14.365$ and $a=0.257$ for
$\alpha=0.2$. Then, we have the critical chemical potential
$\mu_{c}=\mu_{min}$ \cite{Cai-Li-Zhang}. In Table
\ref{CriticalZheng1}, we present the critical chemical potential
$\mu_{c}$ for chosen values of the Gauss-Bonnet coupling $\alpha$
and various masses of the scalar field determined by fixing
$m^{2}L_{\rm eff}^2$. Comparing with numerical results, we find that
the analytic results derived from S-L method are in good agreement
with the numerical calculation.

\begin{table}[ht]
\caption{\label{CriticalZheng1} The critical chemical potential
$\mu_{c}$ obtained by the analytical S-L method (left column) and
from numerical calculation (right column) with chosen Gauss-Bonnet
coupling and various masses of the scalar field for the s-wave
holographic insulator and superconductor model. In order to compare
with the results in Refs. \cite{Nishioka-Ryu-Takayanagi,Pan-Wang},
we also present the critical chemical potential for $m^{2}L_{\rm
eff}^2=-15/4$.}
\begin{tabular}{c c c c }
         \hline
$\alpha$ & 0.0001 & 0.1 & 0.2
        \\
        \hline
~~~~$m^2L_{\rm
eff}^2=0$~~~~&~~~~~$3.407$~~~~~~~$3.404$~~~~~&~~~~~$3.559$~~~~~~~$3.556$~~~~~&
~~~~~$3.790$~~~~~~~$3.789$~~~~~
          \\
~~~~$m^2L_{\rm eff}^2=-1$~~~~&~~~~~$3.137$~~~~~~~$3.135$~~~~~&
~~~~~$3.275$~~~~~~~$3.272$~~~~~&~~~~~$3.477$~~~~~~~$3.475$~~~~~
          \\
~~~~$m^2L_{\rm
eff}^2=-2$~~~~&~~~~~$2.817$~~~~~~~$2.815$~~~~~&~~~~~$2.937$~~~~~~~$2.935$~~~~~&
~~~~~$3.106$~~~~~~~$3.105$~~~~~
          \\
~~~~$m^2L_{\rm
eff}^2=-3$~~~~&~~~~~$2.399$~~~~~~~$2.396$~~~~~&~~~~~$2.497$~~~~~~~$2.494$~~~~~&
~~~~~$2.624$~~~~~~~$2.622$~~~~~
          \\
~~~~$m^2L_{\rm
eff}^2=-15/4$~~~~&~~~~~$1.897$~~~~~~~$1.888$~~~~~&~~~~~$1.963$~~~~~~~$1.960$~~~~~&
~~~~~$2.042$~~~~~~~$2.039$~~~~~
          \\
        \hline
\end{tabular}
\end{table}

From Table \ref{CriticalZheng1}, we observe that for the same mass
of the scalar field, the critical chemical potential increases when
the Gauss-Bonnet factor $\alpha$ becomes bigger. Our analytic result
supports the observation obtained numerically that the higher
curvature correction can make the scalar hair more difficult to be
developed
\cite{Pan-Wang,Gregory,KannoCQG,Brihaye,Ge-Wang,Ge2011,Liu-Wang,BarclayGregory,Pan-WangPLB,Cai-Nie-Zhang,Siani,Setare-Momeni}.
It is interesting to note that the analytical S-L method can give
consistent critical chemical potential with the numerical result
even when the mass of the scalar field is zero. This shows that the
S-L method is more effective than the matching method. It was
realized that the matching method cannot deal with the scalar field
with zero mass \cite{Pan-Wang}, since for this case the Gauss-Bonnet
term does not contribute to the analytic approximation in the
matching method \cite{Gregory,Pan-Wang}.

For the same strength of the curvature correction, with the
increase of the mass of scalar field, the critical chemical potential
$\mu_{c}$  becomes larger. This property also agrees well with the numerical result
\cite{Pan-Wang}.

In the AdS black hole in Gauss-Bonnet gravity \cite{Li-Cai-Zhang},
the S-L method was applied by choosing specific mass of the scalar
field by fixing $m^2L_{eff}^2=-3$. For choosing other nonzero values of
the mass of scalar field, (4.16) there cannot  be integrated
analytically so that their (4.17), (4.18) and further steps cannot
be derived. In the AdS soliton background, we found a much better
situation. (18) above can be  integrated analytically when the mass
of the scalar field satisfying the Breitenlohner-Freedman bound
\cite{Breitenloher}, so that analytically we can observe the
condensation with the change of the mass of the scalar field.

If we fix the scalar field mass by choosing values of $m^{2}L^2$
instead of $m^{2}L_{\rm eff}^2$, the S-L method can give the same
qualitative dependence of the critical chemical potential on the
Gauss-Bonnet factor as described above when we study the scalar
operator $\langle{\cal O}_{+}\rangle$.  This supports the numerical
computation in the AdS black hole background
\cite{Gregory,Pan-Wang}.

Now we concentrate on the scalar operator  $\langle{\cal
O}_{-}\rangle$ by imposing the condition $\psi_{+}=0$. Fixing the
mass of the scalar field by choosing values of $m^{2}L_{\rm eff}^2$,
the scalar operator $\langle{\cal O}_{-}\rangle$ presents us
qualitatively the same behavior of the condensation when we change
the strength of the curvature correction as we observed above. With
the increase of the Gauss-Bonnet coupling, the critical chemical
potential will increase, which shows that the condensation will be
harder to develop. But if we fix the mass of the scalar field by
choosing values of $m^{2}L^2$, analytically we observed completely
different condensation behavior as the Gauss-Bonnet coupling
changes, see Table \ref{CriticalFu}. Our analytical result got by
using S-L method presents the same abnormal behavior as found
numerically in \cite{Pan-Wang}. Considering that choosing the mass
of the scalar field by selecting the value of $m^{2}L_{\rm eff}^2$
contains directly the signature of Gauss-Bonnet factor in the scalar
mass, we believe that this way of choosing the scalar field mass can
disclose the correct consistent influence due to the Gauss-Bonnet
coupling in various condensates.

\begin{table}[ht]
\caption{\label{CriticalFu} The critical chemical potential
$\mu_{c}$ obtained by using the analytical S-L method for the s-wave
holographic insulator and superconductor model.  The mass of scalar
field is chosen by fixing
 $m^2L_{\rm eff}^2=-15/4$ and $m^2L^2=-15/4$, respectively. }
\begin{tabular}{c c c c c }
         \hline
$\alpha$ & 0.0001 & 0.01 & 0.05 & 0.1
        \\
        \hline
~~$m^2L_{\rm eff}^2=-15/4$~~&~~~~~~~$0.837$~~~~~~~&
~~~~~~~$0.839$~~~~~~~&~~~~~~~$0.849$~~~~~~~&~~~~~~~$0.862$~~~~~~~
          \\
        \hline
~~$m^2L^2=-15/4$~~&~~~~~~~$0.836$~~~~~~~&~~~~~~~$0.798$~~~~~~~&
~~~~~~~$0.648$~~~~~~~&~~~~~~~$0.463$~~~~~~~
          \\
          \hline
\end{tabular}
\end{table}

\subsection{Critical phenomena}

We will use the S-L method to analytically discuss the critical
phenomena for the phase transition between the s-wave holographic
insulator and superconductor in the Gauss-Bonnet gravity. We will
concentrate on studying the critical exponent for condensation
operator and the relations between the charge density and the
chemical potential.

The scalar field $\psi$ can be given by Eq. (\ref{PhiFz}) when
$\mu\rightarrow\mu_{c}$, so we can rewrite the equations of motion
(\ref{Phi-Z}) as
\begin{eqnarray}
\phi^{\prime\prime}+\left(\frac{f^\prime}{f}+\frac{1}{z}\right)
\phi^\prime-\frac{2\langle{\cal O}_{i}\rangle^{2}
z^{2\lambda_i-4}F^2}{f}\phi=0.
\end{eqnarray}
Since the condensation for the scalar operator $\langle{\cal
O}_{i}\rangle$ is so small, we can expand $\phi(z)$ in small
$\langle{\cal O}_{i}\rangle$ as
\begin{eqnarray}
\phi(z)\sim\mu_{c}+\langle{\cal O}_{i}\rangle\chi(z)+\cdots.
\end{eqnarray}
Considering the boundary condition at the tip, we can get
$\chi(1)=0$ and $\chi'(1)=constant$. After defining a function
\begin{eqnarray}\label{pzF}
P(z)=\frac{\sqrt{1+4(z^{4}-1)\alpha}-1}{2\sqrt{\alpha}z},
\end{eqnarray}
we can obtain the equation of motion for $\chi(z)$
\begin{eqnarray}\label{chi}
(P\chi')'-2\langle{\cal
O}_{i}\rangle\mu_{c}\frac{z^{2\lambda_i-4}PF^2}{f}=0.
\end{eqnarray}

According to the asymptotic behavior in Eq. (\ref{infinity}), we can
expand $\phi$ when $z\rightarrow0$ as
\begin{eqnarray}\label{expandingPhi}
\phi(z)\simeq\mu-\rho z^{2}\simeq\mu_{c}+\langle{\cal
O}_{i}\rangle[\chi(0)+\chi'(0)z+\frac{1}{2}\chi''(0)z^{2}+\cdots].
\end{eqnarray}
From the coefficients of the $z^{0}$ term, we can easily get
\begin{eqnarray}
\mu-\mu_{c}\simeq\langle{\cal O}_{i}\rangle\chi(0).
\end{eqnarray}
If we set
\begin{eqnarray}
\chi(z)=2\langle{\cal O}_{i}\rangle\mu_{c}\xi(z),
\end{eqnarray}
where the function $\xi(z)$ is the solution to the following
equation
\begin{eqnarray}
\xi^{\prime\prime}+\left(\frac{f^\prime}{f}+\frac{1}{z}\right)
\xi^\prime-\frac{z^{2\lambda_i-4}F^2}{f}=0,
\end{eqnarray}
we will know that
\begin{eqnarray}\label{operatorZF}
\langle{\cal
O}_{i}\rangle=\frac{1}{[2\mu_{c}\xi(0)]^{1/2}}(\mu-\mu_{c})^{1/2},
\end{eqnarray}
where
$\xi(0)=c_{1}-\int_{0}^{1}[c_{2}+\int_{1}^{z}F(x)^{2}x^{2\lambda_i-3}dx]\frac{dz}{zf(z)}$
with the integration constants $c_{1}$ and $c_{2}$ determined by the
boundary condition $\chi(z)$. For example, fixing $m^2L_{\rm
eff}^2=-15/4$ and $\alpha=0.0001$, we can get $\xi(0)=0.0815$ when
$a=0.330$ which results in $\langle{\cal
O}_{+}\rangle\approx1.801(\mu-\mu_{c})^{1/2}$. This agrees well with
the result given in \cite{Cai-Li-Zhang}.

Note that our expression (\ref{operatorZF}) is valid for all cases
considered here, thus near the critical point, both of the scalar
operators $\langle{\cal O}_{+}\rangle$ and $\langle{\cal
O}_{-}\rangle$ satisfy $\langle{\cal
O}_{i}\rangle\sim(\mu-\mu_{c})^{1/2}$. This behavior holds for
various values of Gauss-Bonnet couplings and masses of the scalar
field. The analytic result supports the numerical computation
\cite{Liu-Wang,Pan-Wang} that the phase transition between the
s-wave holographic insulator and superconductor belongs to the
second order and the critical exponent of the system takes the
mean-field value $1/2$. The Gauss-Bonnet coupling will not influence
the result.

Considering the coefficients of  $z^{1}$ terms in Eq.
(\ref{expandingPhi}), we find that $\chi'(0)\rightarrow0$ if
$z\rightarrow0$, which is consistent with the following relation by
integrating both sides of Eq. (\ref{chi})
\begin{eqnarray}\label{chiz0}
\left[\frac{\chi'(z)}{z}\right]\bigg|_{z\rightarrow
0}=-\frac{4\sqrt{\alpha}\langle{\cal
O}_{i}\rangle\mu_{c}}{\sqrt{1-4\alpha}-1}\int_{0}^{1}\frac{z^{2\lambda_i-4}PF^2}{f}dz.
\end{eqnarray}

Comparing the coefficients of the $z^{2}$ term in Eq.
(\ref{expandingPhi}), we can express $\rho$ as
\begin{eqnarray}
\rho=-\frac{1}{2}\langle{\cal O}_{i}\rangle\chi''(0).
\end{eqnarray}
From Eqs. (\ref{chi}) and (\ref{chiz0}), we arrive at
\begin{eqnarray}
\chi''(0)=\left[\frac{P'(z)}{P(z)}\chi'(z)\right]\bigg|_{z\rightarrow
0}=-\frac{4\sqrt{\alpha}\langle{\cal
O}_{i}\rangle\mu_{c}}{\sqrt{1-4\alpha}-1}\int_{0}^{1}\frac{z^{2\lambda_i-4}PF^2}{f}dz.
\end{eqnarray}
Using the above formula and Eq. (\ref{operatorZF}), we can deduce
\begin{eqnarray}
\rho=\Gamma(\alpha,m)(\mu-\mu_{c}),
\end{eqnarray}
where $\Gamma(\alpha,m)$ is a function of the Gauss-Bonnet coupling
and the scalar field mass
\begin{eqnarray}
\Gamma(\alpha,m)=\frac{\sqrt{\alpha}}{(\sqrt{1-4\alpha}-1)\xi(0)}
\int_{0}^{1}\frac{z^{2\lambda_i-4}PF^2}{f}dz.
\end{eqnarray}
Fixing $m^2L_{\rm eff}^2=-15/4$ and $\alpha=0.0001$, for example, we
can get $\Gamma(\alpha,m)=1.330$ when $a=0.330$, so that
$\rho=1.330(\mu-\mu_{c})$ for considering the scalar operator
$\langle{\cal O}_{+}\rangle$.  This is consistent with the result
given in \cite{Cai-Li-Zhang}. Here we observed that the Gauss-Bonnet
coupling will not alter the result. Our analytic finding of a linear
relation between the charge density and the chemical potential
$\rho\sim(\mu-\mu_{c})$ supports the numerical result reported in
\cite{Pan-Wang}.

\section{Phase transition between the p-wave insulator and superconductor}

Since the S-L method is effective to obtain the s-wave holographic
insulator and superconductor phase transition, we will use it to
investigate analytically the p-wave holographic insulator and
superconductor phase transition in the Gauss-Bonnet gravity which
has not been constructed as far as we know.

Considering an $SU(2)$ Yang-Mills action in the bulk theory
\cite{GubserPRL08}
\begin{eqnarray}\label{p-System}
S=\int d^{5}x\sqrt{-g}\left(
-\frac{1}{4}F^{a}_{\mu\nu}F^{a\mu\nu}\right),
\end{eqnarray}
where
$F^{a}_{\mu\nu}=\partial_{\mu}A^{a}_{\nu}-\partial_{\nu}A^{a}_{\mu}+\epsilon^{abc}A^{b}_{\mu}A^{c}_{\nu}$
is the $SU(2)$ Yang-Mills field strength and $\epsilon^{abc}$ is the
totally antisymmetric tensor with $\epsilon^{123}=+1$. The
$A^{a}_{\mu}$ are the components of the mixed-valued gauge fields
$A=A^{a}_{\mu}\tau^{a}dx^{\mu}$, where $\tau^{a}$ are the three
generators of the $SU(2)$ algebra satisfy
$[\tau^{a},\tau^{b}]=\epsilon^{abc}\tau^{c}$.

In order to construct a p-wave holographic insulator and
superconductor in the Gauss-Bonnet gravity, we adopt the ansatz of
the gauge fields as
\cite{Cai-Li-Zhang,Gubser-Pufu,Basu-pwave,Ammon-pwave,Akhavan-Soliton},
\begin{eqnarray}\label{p-ansatz}
A(r)=\phi(r)\tau^{3}dt+\psi(r)\tau^{1}dx.
\end{eqnarray}
Here we regard the $U(1)$ symmetry generated by $\tau^{3}$ as the
$U(1)$ subgroup of $SU(2)$. The gauge boson with nonzero component
$\psi(r)$ along $x$-direction is charged under $A^{3}_{t}=\phi(r)$.
According to AdS/CFT correspondence, $\phi(r)$ and $\psi(r)$ are
dual to the chemical potential and the $x$-component of some charged
vector operator $O$ in the boundary field theory respectively. The
condensation of $\psi(r)$ will spontaneously break the $U(1)$ gauge
symmetry and lead to a phase transition, which can be interpreted as
a p-wave insulator and superconductor phase transition on the
boundary.

From the Yang-Mills action (\ref{p-System}), we can derive the
following equations of motion
\begin{eqnarray}
\psi^{\prime\prime}+\left(
\frac{f^\prime}{f}+\frac{1}{r}\right)\psi^\prime
+\frac{\phi^2}{r^2f}\psi=0, \label{pwave-Psi}
\end{eqnarray}
\begin{eqnarray}
\phi^{\prime\prime}+\left(\frac{f^\prime}{f}+\frac{1}{r}\right)
\phi^\prime-\frac{\psi^2}{r^2f}\phi=0, \label{pwave-Phi}
\end{eqnarray}
where the prime denotes the derivative with respect to $r$.

In order to solve the above equations of motion, we have to impose
the boundary conditions for the $\phi(r)$ and $\psi(r)$ fields at
the tip $r=r_{s}$ and at $r\rightarrow\infty$. At the tip $r=r_{s}$,
the solutions have the same form just as Eq. (\ref{SolitonBoundary})
for the s-wave holographic insulator and superconductor model. But
near the boundary $r\rightarrow\infty$, we have different asymptotic
behaviors
\begin{eqnarray}
\psi=\psi_{0}+\frac{\psi_{2}}{r^{2}}\,,\hspace{0.5cm}
\phi=\mu-\frac{\rho}{r^{2}}\,, \label{p-infinity}
\end{eqnarray}
where $\mu$ and $\rho$ are interpreted as the chemical potential and
charge density in the dual field theory, while $\psi_{0}$ and
$\psi_{2}$ may be identified as a source and the expectation value
of the dual operator, respectively. Since we are interested in the
case where the dual operator is not sourced, we will set
$\psi_{0}=0$ and have a normalizable solution.

\subsection{Critical chemical potential}

Define the variable $z=r_{s}/r$, the equations of motion
(\ref{pwave-Psi}) and (\ref{pwave-Phi}) can be expressed in the z
coordinate as
\begin{eqnarray}
\psi^{\prime\prime}+\left(
\frac{f^\prime}{f}+\frac{1}{z}\right)\psi^\prime
+\frac{\phi^2}{z^2f}\psi=0\,, \label{pwave-PsiZ}
\end{eqnarray}
\begin{eqnarray}
\phi^{\prime\prime}+\left(\frac{f^\prime}{f}+\frac{1}{z}\right)
\phi^\prime-\frac{\psi^2}{z^2f}\phi=0, \label{pwave-PhiZ}
\end{eqnarray}
where the prime denotes the derivative with respective to $z$.

\begin{figure}[H]
\includegraphics[scale=0.75]{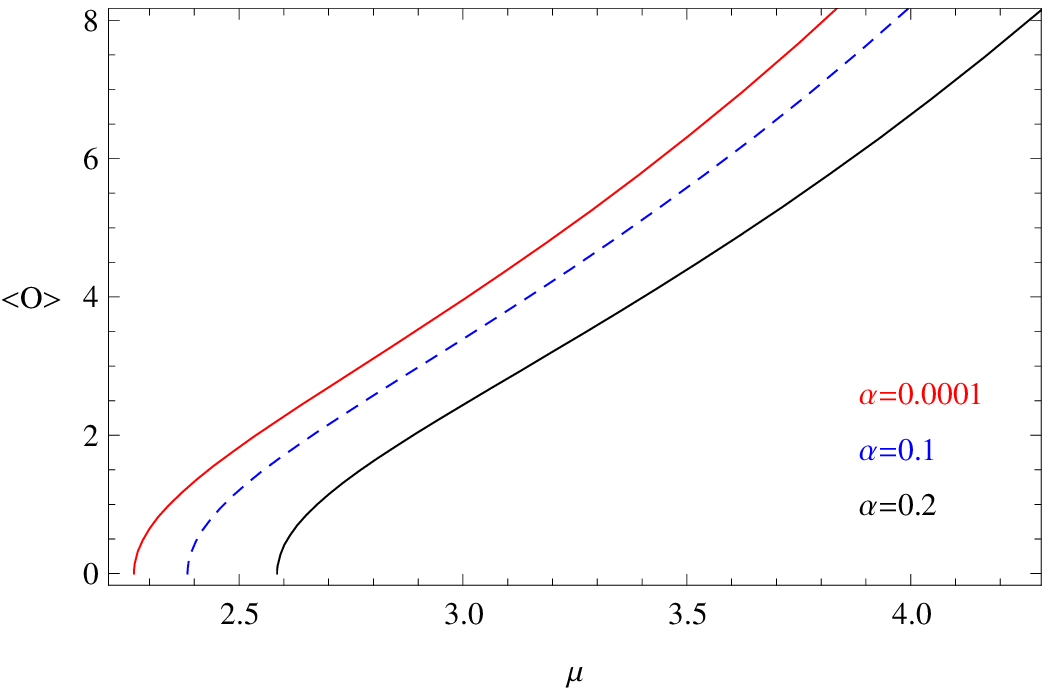}\hspace{0.2cm}%
\includegraphics[scale=0.75]{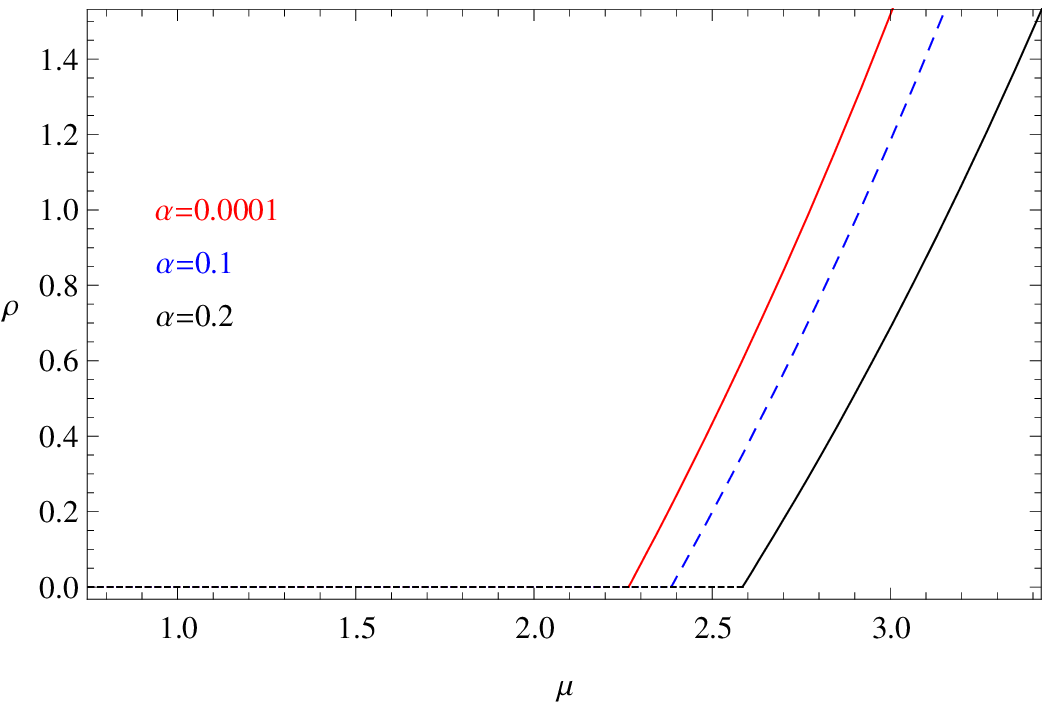}\\ \vspace{0.0cm}
\caption{\label{Cond-Realization} (color online) The condensates of
the operator $\langle{\cal O}\rangle=\psi_{2}$ and charge density
$\rho$ with respect to the chemical potential $\mu$ for different
Gauss-Bonnet couplings $\alpha$ for the p-wave holographic insulator
and superconductor model. The three lines from left to right
correspond to increasing $\alpha$, i.e., $\alpha=0.0001$ (red),
$0.1$ (blue and dashed) and $0.2$ (black) respectively.}
\end{figure}

Similar to the analysis in the previous section, if
$\mu\leq\mu_{c}$, the field $\psi$ is nearly zero, i.e.,
$\psi\simeq0$. Thus, we can obtain the physical solution
$\phi(z)=\mu$ to Eq. (\ref{pwave-PhiZ}) when $\mu<\mu_{c}$. This is
consistent with the numerical results in Fig. \ref{Cond-Realization}
which plot the condensates of the operator $\langle{\cal
O}\rangle=\psi_{2}$ and charge density $\rho$ with respect to the
chemical potential $\mu$ for different Gauss-Bonnet couplings
$\alpha$.

As $\mu\rightarrow\mu_{c}$, Eq. (\ref{pwave-PsiZ}) will become
\begin{eqnarray}
\psi^{\prime\prime}+\left(
\frac{f^\prime}{f}+\frac{1}{z}\right)\psi^\prime
+\frac{\mu^2}{z^2f}\psi=0.
\end{eqnarray}
We can also define a trial function $F(z)$ near the boundary $z=0$
just as in the last section
\begin{eqnarray}\label{pwave-PhiFz}
\psi(z)\sim \langle{\cal O}\rangle z^{2}F(z),
\end{eqnarray}
with the boundary condition $F(0)=1$ and $F'(0)=0$. Therefore, the
equation of motion for $F(z)$ is given by
\begin{eqnarray}\label{pwave-Fzmotion}
F^{\prime\prime}+\left(\frac{5}{z}+\frac{f'}{f}\right)
F^{\prime}+\left(\frac{4}{z^2}+\frac{2f'}{zf}+\frac{\mu^{2}}{z^{2}f}\right)F=0.
\end{eqnarray}
Introducing a new function
\begin{eqnarray}
T(z)=\frac{z^{3}(\sqrt{1+4(z^{4}-1)\alpha}-1)}{2\sqrt{\alpha}},
\end{eqnarray}
we can rewrite Eq. (\ref{pwave-Fzmotion}) as
\begin{eqnarray}\label{pwave-NewFzmotion}
(TF^{\prime})^{\prime}+T\left(\frac{4}{z^2}+\frac{2f'}{zf}+\frac{\mu^{2}}{z^{2}f}\right)F=0.
\end{eqnarray}
Defining the following parameters
\begin{eqnarray}
U=\frac{4}{z^2}+\frac{2f'}{zf},~~V=\frac{T}{z^{2}f},
\end{eqnarray}
we find that, following the Sturm-Liouville eigenvlaue problem
\cite{Gelfand-Fomin}, the minimum eigenvalue of $\mu^2$ can be
obtained from variation of the following functional
\begin{eqnarray}\label{pwaveeigenvalue}
&&\mu^{2}=\frac{\int^{1}_{0}T\left(F'^{2}-UF^{2}\right)dz}{\int^{1}_{0}VF^{2}dz}\nonumber\\
&&\quad~=\frac{1}{4[6+(3a-8)a]\alpha^{5/2}}\left\{4\sqrt{\alpha}\left[12\alpha(1-4a)+a(8-3a+20a\alpha)-8a(1-4\alpha)^{3/2}\right]\right.\nonumber\\
&&\quad\quad\quad\left.+3(1-4\alpha)[4\alpha+a^{2}(4\alpha-1)]\log\left[\frac{(1-4\alpha)\alpha}{(2\alpha+\sqrt{\alpha})^{2}}\right]\right\},
\end{eqnarray}
where we have assumed the trial function to be $F(z)=1-az^{2}$ with
a constant $a$.

We can easily obtain the minimum eigenvalues of $\mu^{2}$ and the
corresponding values of $a$ for different Gauss-Bonnet couplings
$\alpha$, for example, $\mu_{min}^{2}=5.140$ and $a=0.338$ for
$\alpha=0.0001$, $\mu_{min}^{2}=5.697$ and $a=0.305$ for
$\alpha=0.1$ and $\mu_{min}^{2}=6.691$ and $a=0.223$ for
$\alpha=0.2$. Thus, we get the critical chemical potential
$\mu_{c}=\mu_{min}$ \cite{Cai-Li-Zhang} which has been shown in
Table \ref{pwaveCritical} for fixed value of the Gauss-Bonnet factor
$\alpha$. In order to compare with numerical results, we also give
the critical chemical potential obtained by using the shooting
method. Obviously, the agreement of the analytic results derived
from S-L method with the numerical calculation is quite impressive.

\begin{table}[ht]
\caption{\label{pwaveCritical} The critical chemical potential
$\mu_{c}$ obtained by the analytical S-L method (left column) and
from numerical calculation (right column) with fixed Gauss-Bonnet
coupling for the p-wave holographic insulator and superconductor
model. Note that our result reduces to the result in Ref.
\cite{Akhavan-Soliton} if $\alpha\rightarrow0$.}
\begin{tabular}{c c c c }
         \hline
$\alpha$ & 0.0001 & 0.1 & 0.2
        \\
        \hline
$\mu_{c}$&~~~~~$2.267$~~~~~~~$2.265$~~~~~&~~~~~$2.387$~~~~~~~$2.385$~~~~~&
~~~~~$2.587$~~~~~~~$2.585$~~~~~
          \\
        \hline
\end{tabular}
\end{table}

From Table \ref{pwaveCritical}, we also find that the critical
chemical potential increases when the Gauss-Bonnet factor $\alpha$
becomes bigger, which shows that the higher order curvature
corrections in general make the condensation harder to form, just as
observed for the s-wave holographic insulator and superconductor
model. This property agrees well with the numerical result shown in
Fig. \ref{Cond-Realization}.

\subsection{Critical phenomena}

With Eq. (\ref{pwave-PhiFz}), when $\mu\rightarrow\mu_{c}$ the
equation of motion (\ref{pwave-PhiZ}) can be rewrited as
\begin{eqnarray}
\phi^{\prime\prime}+\left(\frac{f^\prime}{f}+\frac{1}{z}\right)
\phi^\prime-\frac{\langle{\cal O}\rangle^{2} z^{2}F^2}{f}\phi=0.
\end{eqnarray}
Note that the condensation value of $\psi(z)$ is so small, we will
expand $\phi(z)$ in small $\langle{\cal O}\rangle$ as
\begin{eqnarray}
\phi(z)\sim\mu_{c}+\langle{\cal O}\rangle\chi(z)+\cdots,
\end{eqnarray}
with the boundary condition $\chi(1)=0$ at the tip. Using the
function defined in Eq. (\ref{pzF}), we can get the equation of
motion for $\chi(z)$
\begin{eqnarray}\label{pwavechi}
(P\chi')'-\langle{\cal O}\rangle\mu_{c}\frac{z^{2}PF^2}{f}=0.
\end{eqnarray}

Near $z\rightarrow0$, we can also expand $\phi$ as
\begin{eqnarray}\label{pwaveEXPPhi}
\phi(z)\simeq\mu-\rho z^{2}\simeq\mu_{c}+\langle{\cal
O}\rangle[\chi(0)+\chi'(0)z+\frac{1}{2}\chi''(0)z^{2}+\cdots].
\end{eqnarray}
Comparing the coefficients of the $z^{0}$ term in both sides of the
above formula, we can obtain
\begin{eqnarray}
\mu-\mu_{c}\simeq\langle{\cal O}\rangle\chi(0).
\end{eqnarray}
Considering the following equation for $\xi(z)$
\begin{eqnarray}
\xi^{\prime\prime}+\left(\frac{f^\prime}{f}+\frac{1}{z}\right)
\xi^\prime-\frac{z^{2}F^2}{f}=0,
\end{eqnarray}
with
\begin{eqnarray}
\chi(z)=\langle{\cal O}\rangle\mu_{c}\xi(z),
\end{eqnarray}
we will have
\begin{eqnarray}\label{pwaveoperator}
\langle{\cal
O}\rangle=\frac{1}{[\mu_{c}\xi(0)]^{1/2}}(\mu-\mu_{c})^{1/2},
\end{eqnarray}
where
$\xi(0)=c_{1}-\int_{0}^{1}[c_{2}+\int_{1}^{z}F(x)^{2}x^{3}dx]\frac{dz}{zf(z)}$
with the integration constants $c_{1}$ and $c_{2}$ determined by the
boundary condition $\chi(z)$. For example, for the case of
$\alpha=0.0001$, we can get $\xi(0)=0.0673$ when $a=0.338$ which
results in $\langle{\cal O}\rangle\approx2.560(\mu-\mu_{c})^{1/2}$.
This is in good agreement with the result given in
\cite{Cai-Li-Zhang,Akhavan-Soliton}.

It should be noted that the relation (\ref{pwaveoperator}) is valid
for all cases considered here, so the condensation $\langle{\cal
O}\rangle\sim(\mu-\mu_{c})^{1/2}$ near the critical point for
various values of Gauss-Bonnet couplings, which agrees well the
numerical results in Fig. \ref{Cond-Realization} that the phase
transition between the p-wave holographic insulator and
superconductor belongs to the second order and the critical exponent
of the system takes the mean-field value $1/2$.

From the coefficients of the $z^{1}$ term in Eq.
(\ref{pwaveEXPPhi}), we obtain that $\chi'(0)\rightarrow0$ which is
consistent with the following relation by making integration of both
sides of Eq. (\ref{pwavechi})
\begin{eqnarray}\label{pwavechiz0}
\left[\frac{\chi'(z)}{z}\right]\bigg|_{z\rightarrow
0}=-\frac{2\sqrt{\alpha}\langle{\cal
O}\rangle\mu_{c}}{\sqrt{1-4\alpha}-1}\int_{0}^{1}\frac{z^{2}PF^2}{f}dz.
\end{eqnarray}

For the coefficients of the $z^{2}$ term in Eq. (\ref{pwaveEXPPhi}),
we have
\begin{eqnarray}
\rho=-\frac{1}{2}\langle{\cal
O}\rangle\chi''(0)=\Gamma(\alpha)(\mu-\mu_{c}),
\end{eqnarray}
where $\Gamma(\alpha)$ is only the function of the Gauss-Bonnet
couplings which can be given by
\begin{eqnarray}
\Gamma(\alpha)=\frac{\sqrt{\alpha}}{(\sqrt{1-4\alpha}-1)\xi(0)}\int_{0}^{1}
\frac{z^{2}PF^2}{f}dz.
\end{eqnarray}
For example, we can obtain $\Gamma(\alpha)=1.126$ when $a=0.338$ for
$\alpha=0.0001$, i.e., the linear relation
$\rho=1.126(\mu-\mu_{c})$, which agrees well with the result given
in \cite{Cai-Li-Zhang}. Still we notice that the Gauss-Bonnet
coupling will not change the result. The analytic finding of a
linear relation between the charge density and the chemical
potential $\rho\sim(\mu-\mu_{c})$ is consistent with the numerical
result presented in Fig. \ref{Cond-Realization}.

\section{conclusions}

We have applied the S-L method to investigate analytically the
condensation and critical phenomena of the phase transition between
the holographic insulator and superconductor in the Gauss-Bonnet
gravity.  We found that unlike the analytic matching method, the S-L
method is effective to obtain the analytic results in the AdS
soliton background both for s-wave (the scalar field) and p-wave
(the vector field) models. Different from the AdS black hole in the
Gauss-Bonnet gravity, in the AdS soliton spacetime we observed that
the S-L method can bring us results of condensation for different
values of the scalar field mass satisfying the
Breitenlohner-Freedman bound. For the massless scalar field, the
information of the Gauss-Bonnet coupling can still be kept in the
S-L method. With this analytic method, we also found that it is more
appropriate to choose the mass of the scalar field by selecting the
value of $m^2L_{\rm eff}^2$. The analytic results derived from the
S-L method for the s-wave and p-wave holographic
insulator/superconductor phase transitions support the numerical
computations and show that the higher curvature corrections make it
harder for the condensation to form.

Furthermore, comparing with the matching method, we found that in
the AdS soliton in the Gauss-Bonnet gravity the S-L method can
present us analytic results on critical exponent of condensation
operator and the relation between the charge density and the
chemical potential near the phase transition point. We observed that
effect of the Gauss-Bonnet factor cannot modify the critical
phenomena. The analytic results can be used to back up the numerical
findings in both s-wave and p-wave insulator/superconductor models
of the Gauss-Bonnet gravity.

\begin{acknowledgments}

This work was supported by the National Natural Science Foundation
of China; the National Basic Research of China under Grant No.
2010CB833004, PCSIRT under Grant No. IRT0964, the Construct Program
of the National Key Discipline, and Hunan Provincial Natural Science
Foundation of China 11JJ7001.

\end{acknowledgments}

\end{document}